\documentclass[
  showpacs,floatfix,
  twoside,
  twocolumn,
  aps,
  prlsds
]{revtex4}

\usepackage[ngerman,american]{babel}
\usepackage{graphicx}
\usepackage{amsmath}
\usepackage{pxfonts}
\usepackage{wasysym}
\usepackage{manfnt}
\usepackage{color}
\usepackage{dingbat}
\usepackage{multirow}

\begin{document}

\title{Universal features of cluster numbers in percolation}
\author{Stephan Mertens$^1$, Iwan Jensen$^2$, and Robert M. Ziff$^3$}
\affiliation{$^1$ Institut f\"ur Theoretische Physik, Otto-von-Guericke Universit\"at, PF 4120, 39016 Magdeburg, Germany\\
    Santa Fe Institute, 1399 Hyde Park Rd., Santa Fe, NM 87501, USA \\  $^2$ School of Mathematics \&\ Statistics, University of Melbourne, Victoria
3010, Australia\\
$^3$ Center for the Study of Complex Systems and Department of Chemical Engineering, University of Michigan, Ann Arbor, Michigan 48109-2136, USA}

\begin{abstract}
The number of clusters per site $n(p)$ in percolation at the critical point $p = p_c$ is not itself a universal quantity---it depends upon the lattice and percolation type (site or bond).  However, many of its properties, including finite-size corrections, scaling behavior with $p$, and amplitude ratios, show various degrees of universal behavior.  Some of these are universal in the sense that the behavior depends upon the shape of the system, but not lattice type.  Here, we elucidate the various levels of universality for elements of $n(p)$ both theoretically and by carrying out extensive studies
on several two- and three-dimensional systems, by high-order series
analysis, Monte-Carlo simulation, and exact enumeration.  We find many new results,
including precise values for $n(p_c)$ for several systems, a clear demonstration of
the singularity in $n''(p)$, and metric scale factors. 
We make use of the matching polynomial of Sykes and Essam to find exact relations 
between properties for lattices and matching lattices. We propose a criterion for an absolute metric factor $b$ based upon the singular behavior of the scaling function, rather than a relative definition of the metric that has previously been used.
\end{abstract}



\maketitle

\section{Introduction}
\label{sec:introduction}

Percolation is the study of connectivity in random systems, particularly of the transition that occurs when the connectivity first becomes long-ranged \cite{stauffer:aharony}.  Examples are the formation of gels in polymer systems \cite{flory:41}, conductivity in random conductor/insulator mixtures \cite{ottavi:clerc:giraud:roussenq:guyon:mitescu:78}, and flow of fluids in random porous materials \cite{larson:scriven:davis:81}.  The percolation model has been of immense theoretical interest in the field of statistical mechanics, being a particularly simple example of a system that undergoes a non-trivial phase transition.  It is directly related to the Ising model through the Fortuin-Kasteleyn \cite{fortuin:kasteleyn:72} representation of the Potts model.  Variations  that have received attention recently  including $k$-core or bootstrap percolation \cite{dorogovtsev:goltsev:mendes:06}, invasion percolation and watersheds \cite{knecht:trump:benavraham:ziff:12,araujo:andrade:ziff:herrmann:11}, and explosive percolation \cite{achlioptas:dsouza:spencer:09,araujo:andrade:ziff:herrmann:11}.  Percolation has also been intensely studied in the mathematical field in recent years \cite{smirnov:werner:01,schramm:smirnov:garban:11,flores:kleban:ziff:11}.  


In the basic model of random percolation, one considers a lattice of sites (vertices) and bonds (edges), and one randomly occupies a fraction $p$ of either sites or bonds, creating clusters of connected components.  Of particular interest is the behavior near the critical threshold $p_c$ where an infinite cluster first appears.
The study of this model has encompassed a wide  variety of approaches, including  experimental measurements \cite{ottavi:clerc:giraud:roussenq:guyon:mitescu:78},  asymptotic analysis of exact series expansions \cite{domb:pearce:76}, theoretical methods  \cite{temperley:lieb:71}, conformal invariance \cite{cardy:92}, Schramm Loewner Evolution theory \cite{smirnov:werner:01,schramm:smirnov:garban:11}, and numerous types of computer simulation  \cite{vyssotsky:etal:61,dean:bird:67,reynolds:stanley:klein:80,tiggemann:01,xu:wang:lv:deng:14, leath:76,hoshen:kopelman:76,newman:ziff:00,newman:ziff:01}.  For some classes of 2d models, thresholds can be found exactly \cite{sykes:essam:64,scullard:ziff:06,grimmett:manolescu:14}, and recently methods have been developed to find approximate 2d values to extremely high precision \cite{scullard:jacobsen:12,jacobsen:14,yang:zhou:li:13,jacobsen:15}.

Universality has played a central role in the understanding of the critical behavior of the percolation process (and in statistical mechanics in general).  First of all there are universal exponents such as $\alpha$ (related to the number of clusters), $\beta$ (the percolation probability $P_\infty$), $\sigma$ (the inverse of the exponent for the divergence of the typical cluster size), $\nu$ (the correlation length) etc.\cite{stauffer:aharony}.  For all systems of a given dimensionality, these exponents have universal values, such as $\alpha = 2/3$, $\beta = 5/36$, $\sigma = 36/91 $  and $\nu = 4/3$ in two dimensions (2d), independent of the system (lattice, non-lattice, etc) and the shape of the boundary.  This is the strongest form of universality.

Secondly, there are quantities, such as the number of clusters of size $s$, $n_s \sim s^{-\tau} f_1(b (p-p_c) s^\sigma)$, whose the scaling function $f_1(z)$ is universal, identical for all systems of a given dimensionality, although in order for this universality to be realized, the metric factor $b$ must be adjusted for each system.  One usually assumes $b = 1$ for one system, such as bond percolation on the square lattice, and then chooses $b$ for the other systems to get the behaviors to match.  The metric factor compensates for the roles of $L$ and $p$ for the different systems.  Here the system is assumed to be infinite, and the scaling function $f_1(z)$ is independent of the system shape that was used in the limiting process to infinity.

Thirdly, there are properties that are universal in the sense of being independent of the lattice and percolation type, but still dependent upon the shape of the system, even in the limit that the system size becomes infinite.  For example, the finite-size scaling of $P_\infty$ is given by
\begin{equation}
    P_\infty(p,L) \sim L^{-\beta/\nu} f_2(b (p-p_c) L^{1/\nu})
\end{equation}
where the scaling function $f_2(z)$ is universal only when comparing different systems of the same shape and boundary condition. (Again, $b$ has to be adjusted to make the different systems coincide, and will be the same $b$ as in $f_1(b(p-p_c)s^\sigma)$.)   The reason that shape matters here is that, for $p$ close to $p_c$, the correlation length diverges, and the boundaries of the system are seen. Note $P_\infty = s_\mathrm{max}/L^d$ is just the size of the maximum cluster divided by the area or volume of the system, and the properties of the maximum cluster will depend upon the boundary of a system.  Another well-known example of a shape-dependent quantity is the percolation crossing probability, where for a rectangular system Cardy derived his well-known formula for the crossing of a rectangular system of any aspect ratio.\cite{cardy:92}.  Here the system is made infinite but with the boundary shape fixed in the limiting process.

The reference to system shape may seem irrelevant, since usually percolation is related to just connectivity.  However, there are finite-size effects that depend upon the large clusters of a system, and for those clusters there is a unique representation of a lattice in space that makes the cluster growth isotropically.  For example, the triangular lattice can be deformed into a square lattice with diagonals in one directions, but in that representation the clusters would grow unequally in the two diagonal directions.  To properly characterize the shape of the system, the triangles must be represented equilaterally.

One of the earliest and most fundamental quantities to be studied in percolation is simply the number of clusters per site $n(p)$ as a function of the occupation probability $p$ \cite{fisher:essam:61,sykes:essam:64}; this quantity corresponds to the free energy of the percolating system \cite{fortuin:kasteleyn:72}. 
In an infinite system and for $p$ near $p_c$, $n(p)$ behaves as
\begin{equation}
  \label{eq:rho-singularity}
  n(p) = A_0 + B_0(p-p_c)+C_0 (p-p_c)^2 +   {\mathcal A}^\pm |p-p_c|^{2-\alpha}+\ldots\,.
\end{equation}
where the first three terms represent the analytical part af $n(p)$, and the last term represents the singular part.  $ {\mathcal A}^\pm$ is the amplitude above ($+$) and below ($-$) the critical point $p_c$.  In two dimensions, the critical exponent $\alpha$ has the universal value $\alpha = -2/3$ \cite{domb:pearce:76} and ${\mathcal A}^+ = {\mathcal A}^-$.  However, the value of ${\mathcal A}^\pm$, as well as those of $A_0$, $B_0$ and $C_0$, are nonuniversal.   The subscript $0$ indicates an infinite system.  The singularity is a weak one and $n(p)$ becomes infinite at $p_c$ in the third derivative.  In terms of the correlation length 
$\xi\sim |p-p_c|^{-\nu}$ where $d \nu = 2 - \alpha$, the singularity in $n(p)$ is proportional to $\xi^{-d}$, where $d$ is the number of dimensions.

 In 1976, Domb and Pearce \cite{domb:pearce:76}, using series analysis, found values of the coefficients $A_0$, $B_0$, $C_0$ and $ {\mathcal A}^\pm$ for two systems: site percolation on the triangular lattice, and bond percolation on the square lattice (see Table \ref{tab:ABC}).  They used their results to conjecture that $\alpha = -2/3$, which proved correct.  However, there has been little further determination or discussion of these 
quantities, other than $A_0$, since then.  One exception is the finite-size correction to $n(p_c)$, the so-called excess cluster number \cite{ziff:finch:adamchik:97}, where measurements have been made and the shape dependence has been quantified theoretically. However, other correction quantities, and especially the strength of the singularity, have not been studied.

In the present paper, we report several new high-precision results for the quantities in (\ref{eq:rho-singularity}), and also discuss, for the first time we believe, many aspects of the finite-size scaling corrections, with a focus on universality.  We determine the metric factors $b$ using the same convention as Hu et al., that $b = 1$ for bond percolation on the square lattice, but then also propose an ``absolute" definite of $b$ by using a fully universal property of the scaling function---the coefficient of the singular behavior, which we can take as equal to unity.  We determine this absolute $b$ for site percolation on the triangular, square, honeycomb, and union-jack lattices, and for bond percolation on the square lattice, where $b$ is no longer equal to 1. 


\section{Finite-size corrections and scaling theory}

The leading amplitude $A_0$ in (\ref{eq:rho-singularity}) gives the critical number of clusters per site $n(p_c)$, and has been found exactly in only two cases: bond percolation on the square lattice, where 
the number of clusters per bond is  \cite{temperley:lieb:71,ziff:finch:adamchik:97}
\begin{equation}
n(p_c) = A_0 = \frac{24 \sqrt{3} - 41}{32} = 0.017788106\ldots
\label{A0squarebond}
\end{equation}
and bond percolation on the dual triangular and honeycomb lattices,
where $n(p_c) = (1/3) [35/4 - 3/p_c^\mathrm{TR} -(1-p_c^\mathrm{TR})^6] = 0.01150783\ldots$ and $n(p_c) = (1/3) [35/4 - 3/p_c^\mathrm{TR} - (p_c^\mathrm{TR})^3] = 0.02331840\ldots$ bond clusters per bond, respectively, with $p_c^\mathrm{TR} = 2 \sin \pi/18$ \cite{baxter:temperley:ashley:78,ziff:finch:adamchik:97}.

The next amplitude $B_0$ is  known exactly for some systems.  
Sykes and Essam \cite{sykes:essam:64} showed that for site percolation  on infinite planar lattices,
\begin{equation}
n(p) - \tilde n(1-p) = \phi(p)
\label{eq:SykesEssamMatching}
\end{equation}
where $\tilde n$ represents the number of clusters on the
\emph{matching lattice} in which the vertices in every face of the
original lattice are completely connected, and $\phi(p)$ is the \emph{matching polynomial} or Euler characteristic \cite{neher:mecke:wagner:08} corresponding to the specific lattice.   For all fully triangulated lattices, such as the triangular and union-jack lattices, as well as the square-bond covering lattice, the matching lattice is identical to the original lattice, $p_c = 1/2$, and $\phi(p) = p - 3 p^2 + 2 p^3$ \cite{sykes:essam:64}, implying\begin{equation}
n'(p_c) = B_0 = \phi'(1/2)/2 = -1/4\,.
\end{equation}
For other lattices, we can find exact results if we include the matching lattice.  For example, for a square (SQ) lattice (site percolation), $\phi(p) = p - 2 p^2 + p^4$, and it follows from (\ref{eq:SykesEssamMatching}) that the following  combinations of quantities are known exactly in terms of $p_c$:
\begin{equation}
\begin{aligned}
A_0^\mathrm{SQ}-A_0^\mathrm{NNSQ} &=  p_c - 2 p_c^2 + p_c^4 = 0.01349562262604(1), \\
B_0^\mathrm{SQ}+B_0^\mathrm{NNSQ} &=  1 - 4 p_c + 4 p_c^3 = -0.537943928141750(5),\\
C_0^\mathrm{SQ}-C_0^\mathrm{NNSQ} &=   - 4  + 12 p_c^2 = 0.2161745687555(3)
\end{aligned}
\label{eq:ABCmatching}
\end{equation}
using $p_c$ from \cite{jacobsen:15}, where NNSQ represents the square lattice with next-nearest-neighbor connections, which is the matching lattice of the square lattice.

Next we consider the behavior for finite systems.  For systems of length scale $L$, (\ref{eq:rho-singularity}) is replaced by \cite{aharony:stauffer:97}
\begin{equation}
  \label{eq:rho-finite}
  n_L(p) = A_0 + B_0(p-p_c)+C_0 (p-p_c)^2 +L^{-d} f(z)+\ldots\,, 
\end{equation}
where $f(z)$ is the leading scaling function. Here $z = b (p-p_c)L^{1/\nu}$ and $b$ is a metric factor depending on the lattice and 
percolation type, but not on the shape of the boundary of the system.  The subscript $L$ on $n_L(p)$ indicates a finite system. We assume that the boundary conditions are periodic, so there are no surface correction terms.  We do not consider higher-order corrections-to-scaling terms, such as $ L^{-2d} g(z)$, here.

The scaling function $f(z)$ depends upon the system's
shape, boundary conditions and dimensionality, but is universal for all percolation types, including different lattices with site or bond percolation, continuum systems, etc., for systems of the same shape.  It is analytic around the origin, allowing us make a Taylor expansion about $z = 0$:
\begin{equation}
\label{eq:nfinite}
n_{L}(p) \sim A + B(p-p_c) + C(p-p_c)^2 + \ldots
\end{equation}
with
\begin{subequations}
\label{eq:ABC-scaling}
\begin{align}
\label{eq:ABC-scaling-a}
  A &= A_0 + A_1 L^{-d} + \ldots   \\ 
  \label{eq:ABC-scaling-b}
   B &= B_0 + B_1 L^{-d+{1}/{\nu}} + \ldots\\
    \label{eq:ABC-scaling-c}
    C &= C_0 + C_1 L^{-d+{2}/{\nu}}+ \ldots
\end{align}
\end{subequations}
and $A_1 = f(0)$, $B_1 = b f'(0)$, $C_1 = b^2 f''(0)/2$. 
The  metric factor $b$ cancels out in the dimensionless ratio
\begin{equation}
R = \frac{A_1 C_1}{B_1^2} = \frac{f(0) f''(0)}{2 f'(0)^2}
\label{eq:R}
\end{equation}
which is predicted to be universal for systems of a given shape.  By including $A_1$ in this ratio, we also account for different definitions of the unit area of the system in $n(p)$, such as using clusters per bond rather than per site for the square-bond system.

For $|z| \gg 1$, $f(z) \sim \hat{\mathcal  A^\pm}  |z|^{2 - \alpha}$,
where $2 - \alpha = d \nu$, $\nu = 4/3$ in 2d and  0.8762 \cite{xu:wang:lv:deng:14} in 3d, and
the amplitudes $\hat {\mathcal  A^\pm}$ are universal for a given definition of $f(z)$.  For large $z$, the behavior is  not shape dependent, because $z \propto (L/\xi)^{1/\nu}$ so for $|z| \gg 1$, $\xi \ll L$ and the boundaries are not seen.
Substituting $z = b(p-p_c)L^{1/\nu}$ into $f(z)$, we find for $z \gg 1$ that $L^{-d} f(z) \sim \hat{\mathcal  A^\pm} b^{2-\alpha}|p-p_c|^{2-\alpha}$, which implies
the singular term in (\ref{eq:rho-singularity})
with 
\begin{equation}
\label{eq:universalA}
{\mathcal A}^\pm = b^{2 - \alpha} \hat {\mathcal  A^\pm}
\end{equation}
This equation shows the scaling between the universal ($\hat {\mathcal  A^\pm}$) and non-universal coefficients (${\mathcal A}^\pm  $) for the different systems.  Note that this implies $B_1/(- {\mathcal  A^\pm})^{1/(2-\alpha)}$ is another universal ratio along with $R$.  We discuss these universal ratios below.

 The correction term $A_1$ in (\ref{eq:ABC-scaling}) is the excess cluster number \cite{ziff:finch:adamchik:97}.  It is the difference between the actual cluster number $L^2 n(p_c)$ and the expected number
$L^2 A_0$, and for compact shapes it is of order 1.  Using results from conformal field theory, $A_1$ can be calculated exactly \cite{ziff:lorenz:kleban:99}, with $A_1=0.883576308\ldots$ for a square torus, and $0.878290117\ldots$ for a 60$^\circ$ periodic rhombus \cite{mertens:jensen:ziff:16}, which is equivalent to a rectangle of aspect ratio $\sqrt{3}/2$ with a twist of $1/2$.  This rhombus is a natural system boundary shape for triangular, hexagonal, and related systems and is conjectured to give the lowest value of $A_1$ for any repeatable shape of a periodic system \cite{ziff:lorenz:kleban:99}.

\begin{table}[htbp]
\caption{
Values of the coefficients $A_0$, $B_0$, $C_0$, $A_1$, $B_1$, and $C_1$ in (\ref{eq:nfinite}) and (\ref{eq:ABC-scaling})
for 2D and 3D systems found in previous papers as cited, and in this work by 
microcanonical MC simulations ($m$), series analysis ($s$), 
conformal field invariance ($c$), or duality ($d$).  Numbers in parentheses give errors in the last digit(s).  
All are for site percolation except
for the square-bond case.  In the latter case, the results are per bond rather than per site on the lattice, accounting for a factor of two decrease in $A_1$ from the other square-boundary cases SQ and UJ. }
\label{tab:ABC}
 \begin{tabular}{lcll}
\hline \hline
Lattice & \multicolumn{1}{c}{$X$} & \multicolumn{1}{c}{$X_0$} &
 \multicolumn{1}{c}{$X_1$} \\
 \hline
Square
&
 $A$ &$\phantom{-}0.027 598
 1(3)$\textsuperscript{\cite{ziff:finch:adamchik:97}}& $\phantom{-}0.8835(5)$\textsuperscript{\cite{ziff:finch:adamchik:97}}\\
& & $\phantom{-}0.02759791(5)$\textsuperscript{\cite{tiggemann:01}} & $\phantom{-}0.883 576 308...$\textsuperscript{\cite{ziff:lorenz:kleban:99}} \\
 & &  $\phantom{-}0.02759800(5)$\textsuperscript{\cite{hu:bloete:deng:12}} &\\
 & &  $\phantom{-}0.02759803(2)$\textsuperscript{m}& $\phantom{-}0.8834(1)$\textsuperscript{m}\\  %
 & $B$ & $-0.3205738(7)$\textsuperscript{m} &$\phantom{-}0.8708(2)$\textsuperscript{m}\\  %
 & $C$
 &$\phantom{-}1.9669(3)$\textsuperscript{m}&$-3.286(3)$\textsuperscript{m}\\
 \hline
 Honeycomb
 &
$A$ & $\phantom{-}0.03530709(1)$\textsuperscript{m} & $\phantom{-}0.9468(1)$\textsuperscript{m}\\
 & & & $\phantom{-}0.946 883 263\ldots$\textsuperscript{c}\\[0.5ex]
& $B$ & $-0.4109549(6)$\textsuperscript{m} & $\phantom{-}0.8260(1)$\textsuperscript{m}\\ 
& $C$ & $\phantom{-}2.3082(2)$\textsuperscript{m} &
$-3.898(1)$\textsuperscript{m}\\
\hline
 Triangular
& $A$ & $\phantom{-}0.0168(2)$\textsuperscript{\cite{domb:pearce:76}} &\\
& & $\phantom{-}0.017 630(2)$\textsuperscript{\cite{margolina:etal:84}} & \\
& & $\phantom{-}0.017 626(1)$\textsuperscript{\cite{rapaport:86}} & \\
& & $\phantom{-}0.017 625 5 (5)$\textsuperscript{\cite{ziff:finch:adamchik:97} } & $\phantom{-}0.878(1)$\textsuperscript{\cite{ziff:finch:adamchik:97} } \\
& & $\phantom{-}0.017625277(4)$\textsuperscript{m} & $\phantom{-}0.87839(7)$\textsuperscript{m} \\
& & $\phantom{-}0.017625277368(2)$\textsuperscript{s} &
$\phantom{-}0.878290117\ldots$\textsuperscript{c} \\ 
& $B$ &  $-0.2500006(3)$\textsuperscript{m} & $\phantom{-}0.8807(1)$\textsuperscript{m}  \\
& &  $-1/4$\textsuperscript{d} &\\[1ex]
& $C$ & $\phantom{-}1.5(2)$\textsuperscript{\cite{domb:pearce:76}} &  \\
& & $\phantom{-}1.91392(9)$\textsuperscript{m} &
$-3.2909(8)$\textsuperscript{m}\\
& & $\phantom{-}1.91391790(5)$\textsuperscript{s} & \\ 
 \hline
Union-Jack
& $A$ & $\phantom{-}0.025662605(6)$\textsuperscript{m} &
$\phantom{-}0.88345(8)$\textsuperscript{m}\\
& & & $\phantom{-}0.883 576 308...$\textsuperscript{\cite{ziff:lorenz:kleban:99}} \\ 
& $B$ & $-0.2500005(3)$\textsuperscript{m} & $\phantom{-}0.76074(5)$\textsuperscript{m}\\
& & $-1/4$\textsuperscript{d} & \\[0.5ex]
& $C$ & $\phantom{-}1.41334(5)$\textsuperscript{m} &
$-2.5206(3)$\textsuperscript{m}\\
\hline
 Square (bond)
& $A$ & $\phantom{-}0.0173(3)$\textsuperscript{\cite{domb:pearce:76}} & \\
& & $\phantom{-}0.017788096(3)$\textsuperscript{m} &
$\phantom{-}0.44183(1)$\textsuperscript{m}\\
& & $\phantom{-}0.017788106(1)$\textsuperscript{s} &  $\phantom{-}0.441 783 154...$ \textsuperscript{\cite{ziff:lorenz:kleban:99}}\\
& & $\phantom{-}0.01778810567665\ldots$ & \\
& & $\phantom{-} = (24 \sqrt{3} - 41)/32$\textsuperscript{\cite{temperley:lieb:71,ziff:finch:adamchik:97}}& \\ 
& $B$ & $-0.2499995(4)$\textsuperscript{m} & $\phantom{-}0.55504(7)$\textsuperscript{m}\\
& & $-1/4$\textsuperscript{d} & \\ 
& $C$ & $\phantom{-}1.4(3)$\textsuperscript{\cite{domb:pearce:76}} & \\
& & $\phantom{-}1.87706(4)$\textsuperscript{m}&
$-2.6882(5)$\textsuperscript{m} \\
&  & $\phantom{-}1.87714(2)$\textsuperscript{s} & \\
\hline

 Cubic
& $A$ & $\phantom{-}0.0524387(3)$\textsuperscript{\cite{tiggemann:01}}  &  \\
 &   & $\phantom{-}0.052 438
 218(3)$\textsuperscript{\cite{wang:etal:13}}  &
 $\phantom{-}0.6746(3)$\textsuperscript{\cite{wang:etal:13}} \\ 
& & $\phantom{-}0.052438223(3)$\textsuperscript{m} &
$\phantom{-}0.6748(2)$\textsuperscript{m}\\ 
& $B$ & $-0.4107249(5)$\textsuperscript{m} & $\phantom{-}1.7147(4)$\textsuperscript{m}\\ 
& $C$ & $\phantom{-}0.4405(6)$\textsuperscript{m} &
$-1.004(7)$\textsuperscript{m}\\
\hline \hline
 \end{tabular}
 \end{table}

\section{Measurements}

 In order to study these quantities, we carried out extensive studies
 using several different methods.  Details will be given in another paper
 \cite{mertens:jensen:ziff:16}.  Many of the results
 are summarized in Table \ref{tab:ABC}, where previous
 values are also listed.  

First of all, we extended the series analysis of $n(p)$ for the
triangular lattice to 69th order.  In 1976, Domb and Pearce
\cite{domb:pearce:76} used a 19th-order analysis to find
$\alpha = -0.668(4)$,  and
they also found accurate values of $A_0$, $B_0$, $C_0$ and $\mathcal A^\pm$.
Using Domb and Pearce's powerful substitution $u = p(1-p)$ on
$B(u) = \phi(p)/2 + n(p)$ \cite{domb:pearce:76} in our series, we
find the very  precise result.
\begin{equation}
n(p_c) = A_0 = 0.017625277368(2)
\label{triangularsite}
\end{equation}
and also to high accuracy the exponent $\alpha = -0.6666669(4)$, an unusually precise test of a critical exponent.
We also checked the result (\ref{A0squarebond}) for $A_0$ of the square bond lattice, and found agreement using a 72-order series (see Table
\ref{tab:ABC}), although the convergence here was slower than for the
triangular lattice.

Secondly, we found exact results for $n(p)$ for small $L \times L$ systems using
the Newman-Ziff (NZ) method \cite{newman:ziff:01}. The NZ method computes $n(p)$ by occupying the sites (or bonds) one by one in random order. The cluster structure can be updated very efficiently because the changes in the cluster structure are triggered by local events. For exhaustive enumerations, we have to loop over all $2^{L\times L}$ configurations and record the cluster structure for each. If you do this in the obvious fashion (binary counting or gray code), many consecutive configurations differ by many occupied sites. In particular many sites change their status from occupied to empty from one configuration to the next. This is something the NZ method cannot handle, and you need to compute the next cluster structure from the empty lattice. There is, however, a clever way to loop through all $2^{L\times L}$ configurations by adding an occupied site most of the time, while the number of transitions that require a restart grows only like $O(2^L)$. With this method, exact computation of $n_L(p)$ is possible for $L\leq 7$ \cite{mertens:jensen:ziff:16}.
For the square lattice with periodic boundary conditions and $L=3$,
for example, the polynomial is
\begin{equation}
\label{polynomial}
\begin{aligned}
n_3(p) =  &9 p q^8 + 54 p^2 q^7 + 132 p^3 q^6 + 171 p^4 q^5\\
& + 135
p^5q^4 + 84p^6q^3+36p^7q^2+9p^8q +p^9\,,
\end{aligned}
\end{equation}
where $q = 1 - p$.   We
considered several systems with $L$ up to 7, and the resulting
polynomials of order $L^2$ are posted on \cite{mertens:website}.

\begin{figure}[tbp]
  \centering
  \includegraphics[width=\columnwidth]{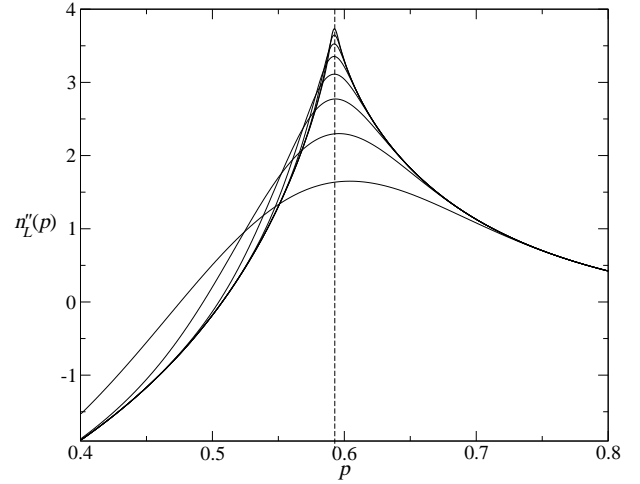}
  \caption{Second derivative of the cluster density $n_L(p)$ for
    square lattices of size $L\times L$ for $L=8,16,\ldots,1024$.  
  Error bars are much smaller than the linewidth. The vertical dashed line marks the percolation threshold $p_c$.}
  \label{fig:2d-example}
\end{figure}

Thirdly, we carried out Monte-Carlo (MC) simulations using the NZ method, which
generates the microcanonical weights---essentially approximations for the coefficients in
polynomials such as (\ref{polynomial}), but for much larger systems.  In this method, occupied sites are added one at a time, and an efficient union-find procedure is used to update the cluster connectivity. 
Once the microcanonical weights $N_{i,L}$ (number of clusters of size $i$ in a system of length $L$) are found, the canonical $p$-dependent expressions are found through a convolution with the binomial distribution:
\begin{equation}
n_L(p) = \frac{1}{N} \sum_{i=0}^N N_{i,L} \binom{N}{i} p^i (1-p)^{N-i}
\label{eq:convolution0}
\end{equation}  
 Derivatives $n^{[k]}_L(p)$ can be found by  a similar convolution 
\begin{equation}
n^{[k]}_L(p) = \frac{1}{N} \sum_{i=0}^N N_{i,L} {\mathcal D_{k,i}} \binom{N}{i} p^i (1-p)^{N-i}
\label{eq:convolution}
\end{equation}  
with ${\mathcal D_{1,i}} = (i - p N)/[p(1-p)]$ and ${\mathcal D_{2,i}}=[i^2 - (1+2(N-1)p) i +N(N-1)p^2]/[p(1-p)]^2$ for the first and second derivatives respectively.

In the MC work we considered $L \times L$ systems with $L$ up to $1024$ for site
percolation (s) on the SQ, NNSQ,
triangular (TR), honeycomb (HC), and union-jack (UJ) lattices, the 3d
cubic lattice, and bond percolation (b) on the SQ lattice.  For the TR lattice, we used a periodic square lattice with diagonal bonds, so the system shape was effectively a 60$^\circ$ rhombus.  For the HC lattice, we also used a square lattice but with half the vertical bonds missing in a brick pattern, so the effective shapes was a rectangle with aspect ratio $\sqrt{3}$.  For each size and lattice type we computed up to 
$10^{10}$ samples.   Fig.\ \ref{fig:2d-example} shows
$n_L''(p)$ for the square-site problem, clearly demonstrating the
development of the branch-point singularity, something not calculated
before.  (Note that peaked plots of closely related 
``specific heat'' functions were given by \cite{kirkpatrick:76} and more recently by \cite{hu:bloete:ziff:deng:14}.)

\begin{figure}
  \centering
  \includegraphics[width=\columnwidth]{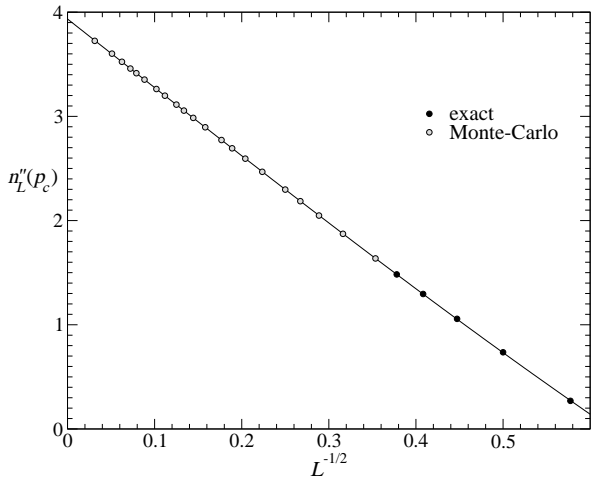}
  \caption{An example of a plot of the MC and exact-enumeration data, used to find
    coefficients given in Table~\ref{tab:ABC}: $n''_{L}(p_c) = 2 C$ for site
    percolation on square lattices vs.\ $L^{-1/2}$. The line is a fit
    of (\ref{eq:ABC-scaling-c}) which yields values for $C_0$ and $C_1$. Error bars of the MC data are much smaller than
    the size of the symbols.}
  \label{fig:extrapolations}
\end{figure}

Finally, we carried out a Monte-Carlo simulation at fixed
$p = p_c$, counting clusters and keeping track of $\langle N_c \rangle$, $\langle N_s \rangle$
and $\langle N_c N_s \rangle$, where $N_c$ is the
number of clusters and $N_s$ is the number of occupied
sites in each sample,  with
$\langle N_s \rangle / L^2 = p$. These allow $B_0 = n'(p_c)$ to be
calculated from
\begin{equation}
n'(p) =\frac{\langle N_s N_c \rangle - \langle N_s \rangle \langle N_c \rangle }{L^2 p(1-p)}
\end{equation}
which follows from (\ref{eq:convolution}) for $k = 1$.
We carried this out for site percolation simultaneously on the matching SQ and NNSQ
lattices, identifying nearest-neighbor clusters on the black sites
(occupied with probability $p$) and next-nearest neighbor clusters on the white
sites (occupied with probability $1-p$) for each sample.  We confirmed our values of $B_0$ and also verified 
 that the matching relation (\ref{eq:ABCmatching}) holds to a high degree of accuracy.
  
Analyzing these results \cite{mertens:jensen:ziff:16}, we find the values of
the  amplitudes listed in Table \ref{tab:ABC}.  Agreement with exact results and with
previous values is generally good.  The early results of Domb and Pearce \cite{domb:pearce:76} have been vastly improved.  Plots of the data of $n_L(p)$, $n_L'(p)$ and $n_L''(p)$ verified that the scaling predicted by (\ref{eq:ABC-scaling}) is correct; for example, the plot for $n_L''(p)$ for site percolation on the square lattice is given in Fig.\ \ref{fig:extrapolations}.

\begin{table}[tpb]
\caption{Values of shape-dependent universal quantities $R = A_1 C_1 / B_1^2$, $B_1/b = B_1/(-{\mathcal A^\pm})^{3/8}$ and $C_2/b^2 = C_2/(-{\mathcal A^\pm})^{3/4}$, using $b[\mathcal A^\pm]$abs.\ from Table \ref{tab:metric}.}
\begin{center}
\begin{tabular}{lcccc}
\hline \hline
System  & Shape & R & $B_1/b$ & $C_1/b^2$\cr
\hline
SQ,b & square & $-3.855(2)$ & $0.4995(1)$ & $-1.0884(2)$ \\
SQ,s & square & $-3.829(4)$ & $0.5002(1)$ & $-1.084(1)$  \\
UJ,s & square & $-3.8484(5)$ & $0.4999(2)$ & $-1.088(1)$  \\
TR,s & rhomb. & $-3.726(2)$ & $0.5064(1)$ & $-1.0883(2)$  \\
HC,s & $\sqrt{3}$ rect. & $-5.410(3)$ & $0.4393(1)$ & $-1.1024(10)$  \\
\hline \hline
\end{tabular}
\end{center}
\label{tab:B1C1square}\
\end{table}%

Calculating the quantity $R$ of (\ref{eq:R}) we find the values given in table \ref{tab:B1C1square}.
The 
three square-boundary
systems give similar values consistent with a common value of
$R=-3.844(10)$, while for TR and HC systems, simulated on a rhombus and rectangle respectively, the value is different.  This confirms our expectations about the shape-dependent but otherwise universal behavior of $R$.

Relative metric factors $b$ can be calculated from $B_1$ and
$C_1$ for systems of the same shape by the equations below (\ref{eq:ABC-scaling}), which imply
\begin{align} 
b/b' &= B_1'/B_1 \label{eq:bB}\\ 
b/b' &= (C_1'/C_1)^{1/2}
\label{eq:bC}
\end{align}
where the prime indicates a reference system.
The relative $b$'s can also be calculated from the $ {\mathcal A}^\pm$, which is not shape-dependent and therefore can be used for all 2d systems we consider, irrespective of the shape that was used in the simulations:
\begin{equation} 
b/b' = [{\mathcal A}^\pm/({\mathcal A}^\pm)']^{3/8}
\label{eq:bA}
\end{equation}
from (\ref{eq:universalA}).  We can choose a convention such as that of Hu et al.\ \cite{hu:lin:chen:95,hu:lin:chen:95b} that $b'=1$ for bond percolation on the square lattice; this yields the values of $b$ given in the first four columns of  Table
\ref{tab:metric}. Note, in order to use this system for a reference, we have to multiply the quantities for the square-bond model by 2 to account for the fact that they represent the number of clusters per bond, not per site, and there are two bonds per site on the square lattice. 

\begin{table}[t]
\caption{Metric factor $b$ calculated from $B_1$ of (\ref{eq:bB}), $C_1$ of (\ref{eq:bC}), and $\mathcal A^\pm$ of (\ref{eq:bA}), normalized to those of the SQ,b system (with a factor of two in the coefficients of the SQ,b system because there are two bonds per lattice site).  Results for $b$ from Hu et al.\ \cite{hu:lin:chen:95,hu:lin:chen:95b} are also shown.  In the last column are the values $b$ based upon the convention $\hat{\mathcal A}^\pm = -1$, calculated from  (\ref{eq:absolute}).}
\begin{center}
\begin{tabular}{lcccc|c}
\hline \hline
Lattice  & $b[B_1]$ & $b[C_1]$ & $b[\mathcal A^\pm]$ & $b$[Hu] & $b[\mathcal A^\pm]$abs.\cr
\hline
SQ,b   & 1& 1 & 1&1 & 2.22254(8) \cr  
SQ,s   & 0.7847(3) & 0.7818(4)  & 0.7810(10) & 0.79& 1.7410(6)\cr
UJ,s  & 0.6854(1) & 0.6847(1) & 0.6815(11) & - & 1.522(6) \cr
TR,s   &  - & - & 0.780(2) & 0.79 & 1.73897548(3) \cr
HC,s & - & - & 0.8435(14) & 0.86 & 1.8804(7)\cr
\hline \hline
\end{tabular}
\end{center}
\label{tab:metric}
\end{table}%

The quantity $\mathcal A^\pm$ can be difficult to measure because, for
a finite-size system, it represents the behavior for sufficiently large $|p - p_c|$
so that $\xi \ll L$, yet still within the scaling region.  Our 2d
results for  $\mathcal A^\pm$  are given in Table \ref{tab:amplitude}.  We also show the
values of $\hat{\mathcal A}^\pm$, the and for the cases
we have measured values of $b$, we find good evidence of universality of that quantity for systems of different shapes.

\begin{table}[htpb]
\caption{The non-universal amplitude $\mathcal A^\pm$ for 2d lattices, with our  series ($s$) and MC ($m$) results, 
along with results from Domb and Pearce \cite{domb:pearce:76}.  The final column shows 
$\hat{\mathcal A}^\pm = b^{-8/3} \mathcal A^\pm$, using our values of $b$
 given in the first two columns of Table~\ref{tab:metric}, representing the SQ,b, SQ,s and UJ systems with the same 
 square boundary.  The results for our measurements on the SQ,b, SQ,s and UJ,s systems gives a fairly 
 consistent value of 8.42. For the last two cases, the HC and TR lattices, we use the values of $b$ 
 from \cite{hu:lin:chen:95} to find $\hat{\mathcal A}^\pm$ from the ${\mathcal A}^\pm$ , 
 and find less consistent values of $\hat{\mathcal A}^\pm$.   For the square-bond system, 
 we have to double the value of $\hat{\mathcal A}^\pm$ because of the different basis used.  
 These values of $\hat{\mathcal A}^\pm$ are based upon the convention that $b = 1$ 
 for bond percolation on the square lattice.
}
\begin{center}
\begin{tabular}{lll}
\hline \hline
Lattice & \quad $-\mathcal A^\pm$ & $-\hat{\mathcal A}^\pm$\cr
\hline
SQ,b & $4.240(15) \textsuperscript{\cite{domb:pearce:76}}$, $4.211(1) \textsuperscript{m}$, 4.2063(2)\textsuperscript{s}, & 8.41\cr
SQ,s  &$ 4.3867(4) \textsuperscript{m}$  & 8.45\cr
UJ,s &  $3.064(3) \textsuperscript{m}$  & 8.40 \cr
TR,s &  $4.370(15) \textsuperscript{\cite{domb:pearce:76}} $,  $4.379(2) \textsuperscript{m}$,  $4.3730310(2) \textsuperscript{s}$   & 8.20 \cr
HC,s & $5.387(5) \textsuperscript{m}$ &  8.05\cr
\hline  \hline
\end{tabular}
\end{center}
\label{tab:amplitude}
\end{table}

\section{Absolute value of the metric factor $b$}

Having verified universality of $\hat{\mathcal A}^\pm$, we can turn it around and 
can use it to propose a definition of $b$ that is not based upon a reference lattice but instead is based upon the universal behavior of $f(z)$.  Because the quantity  $\hat{\mathcal A}^\pm$ is independent of both the lattice type and the system shape, it is a good quantity to use.  There is a freedom to choose an arbitrary overall scale factor for $z$ in $f(z)$, and we can assume that that scale factor is chosen so that $\hat{\mathcal A}^\pm = -1$.  By (\ref{eq:universalA}), this choice implies that $b$ can be calculated from
\begin{equation}
b = (-{\mathcal A}^\pm)^{3/8}
\label{eq:absolute}   \end{equation} 
which leads to the values of $b$ given in the last column of Table \ref{tab:metric}.  We call these ``absolute" values of $b$ because we are not assuming $b=1$ for any particular system.



Using these values for the absolute metric factor $b$, we can find the shape-dependent but otherwise universal behavior of $f(z)$:
\begin{align}
    f(z) &= f(0) + z f'(0) + z^2 f''(0) /2 + \hat{\mathcal{A}^\pm} |z|^{8/3} \\
    &= A_1 + z (B_1/b) + z^2 (C_1/b^2)  - |z|^{8/3} 
\end{align}
For our three systems with the square boundary, we find very good consistency in these coefficients (see Table \ref{tab:B1C1square}) yielding
\begin{equation}
    f(z)= 0.883576 +  0.5000(2)z - 1.088(1) z^2  - |z|^{8/3} \\
    \label{eq:fsquare}
\end{equation}
with the intriguing result that $B_1/b = B_1/(-{\mathcal A}^\pm)^{3/8}$ seems to equal exactly $1/2$ for the square boundary.  We have no explanation for this value.

For the systems with other boundary shapes, we have one system for each.
For a system with a rhombus boundary or equivalently a rectangle of aspect ratio $\sqrt{3}/2$ with a twist of $1/2$ (which we used for the TR lattice), we find
\begin{equation}
    f(z)= 0.878290 +  0.5064(1)z - 1.0883(2) z^2 - |z|^{8/3} \\
        \label{eq:ftr}
\end{equation}
For the HC system, where we used a rectangular boundary of aspect ratio $\sqrt{3}$, we find
\begin{equation}
    f(z)= 0.946883 +  0.4393(1) z -  1.1024(10) z^2  - |z|^{8/3} \\
        \label{eq:fhc}
\end{equation}
Thus, we see, as predicted, that systems of different shapes have different forms of $f(z)$ for small $z$.  Interesting, it seems that  $C_1/b^2$ is the same for the 60$^\circ$ rhombus (the TR system) as for the three square systems.  However, for the $\sqrt{3}$ rectangle (the HC system), it is somewhat different.  We have no explanation for this behavior.

Clearly, an interesting area for future study would be to find $f(z)$ for systems of more shapes, and to also verify universality by considering different lattices of a given shape.

\section{The function $M_L(p)$.}

\begin{figure}[htpb]
  \centering
  \includegraphics[width=\columnwidth]{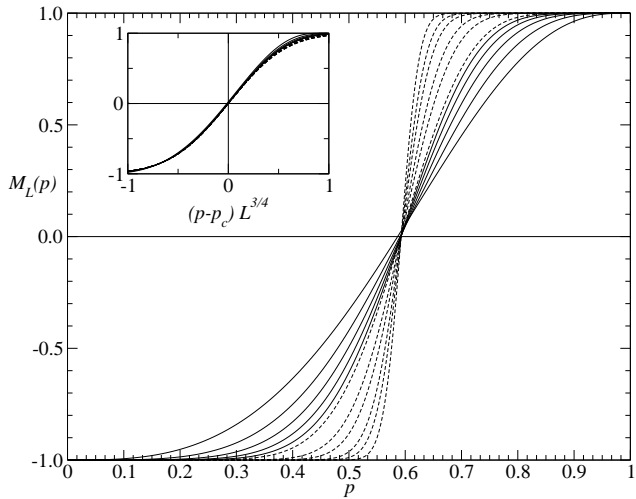}
  \caption{$M_L(p)=L^2[n^\mathrm{SQ}(p) -
    n^\mathrm{NNSQ}(1-p)-\phi(p)]$ vs.\ $p$ from exact enumeration results for $L = 3, 4,
    5, 6, 7$ (solid lines) and $L = 8, 12, 16, 24, 32, 48$ from MC (dashed lines), plotted as a function of $p$, and (inset) as a function of the scaling variable $(p-p_c)L^{1/\nu}$, yielding $f(z)-f(-z)$.  In the limit $L \to \infty$, $M_L(p)$ becomes a step function.}
  \label{fig:matching}
\end{figure}

We also analyzed the function
$M_L(p)=L^2[n_L^{\mathrm{SQ}}(p) - n_L^{\mathrm{NNSQ}}(1-p)-\phi(p)]$,
where $\phi(p) = p - 2 p^2 + p^4$ is the matching polynomial (\ref{eq:SykesEssamMatching}) for the
square lattice. Note that
$M_L(p)/L^2\to0$ as $L\to\infty$, but $M_L(p)$ converges to a step
function independent of $L$ that jumps from $-1$ to $+1$ at $p=p_c$; see
Fig.~\ref{fig:matching}.  At $p_c$, $M_L$ appears to go to zero as 
$M_L(p_c) \sim L^{-4}$ as $L \to \infty$, which implies that finding where  $M_L(p) = 0$ is a very sensitive criterion for finding $p_c$.  In fact, this is identical to the  criterion used by Jacobsen and Scullard  \cite{scullard:jacobsen:12,jacobsen:14,jacobsen:15} whose studies yielded the most precise estimates of percolation thresholds to date.  We discuss $M_L(p)$ more in Ref.\ \cite{mertens:ziff:16}, where it is also shown that $M_L(p)$ is related to the probability of the existence of wrapping clusters on the lattice and matching lattice.

In the inset to Fig.~\ref{fig:matching} we show a plot of $M_L(p)$ as a function of $(p-p_c)L^{1/\nu}$ for the square-site system.  Because of the relations (\ref{eq:ABCmatching}), it follows that in the scaling limit $M_L(p) = f(z) - f(-z)$, all the terms proportional to $L^{2}$ having cancelled out.  If we had plotted the inset to the figure vs.\ $z = b (p - p_c)L^{-1/\nu}$ with $b$ equal to its absolute value $b = 1.741$, then by (\ref{eq:fsquare}) the slope at $z = 0$ would be exactly 1.  

Finally, we also carried out simulations for site percolation on a cubic lattice in three dimensions, and these results are shown in Table \ref{tab:ABC}.  The behavior was found to be consistent with the scaling predictions of equation (\ref{eq:ABC-scaling}).

\section{Conclusions}

In this paper, we have found many new results concerning the function $n(p)$, including 
\begin{itemize}

\item  A discussion of the finite-size corrections to A, B, and C, including a derivation of the scaling of those terms.   

\item  The verification of that scaling on several different system types.

\item A discussion of the use of the coefficient ${\mathcal A}^\pm$ of the singular term in $n(p)$ to define an absolute, rather than relative, value of the metric factor $b$.

\item A visualization of the formation of a cusp in $n''(p)$, Fig.\ \ref{fig:2d-example}.

\item  The extension of previous work on metric factors \cite{hu:lin:chen:95} to a new system, the union-jack lattice.  This system is interesting to study because it is fully triangulated, so has a site threshold of 1/2, but can be made into a perfect square, so is useful to comparing to other square systems.

\item   A discussion of shape-dependent universality \cite{ziff:lorenz:kleban:99,aharony:stauffer:97}, as summarized in Table~\ref{tab:universality}.

\smallskip

\item  Application of the Sykes-Essam matching polynomial to find relations for $A$, $B$, and $C$ between a lattice and its matching lattice.

\item  Development of new algorithms for carrying out the simulations and series analyses.

\item  The determination of many precise values concerning $n(p)$, including a very precise determination of $n(p_c)$ for site percolation on the triangular lattice, using a much extended series expansion for that system.

\item  A discussion of $M_L(z)$ which directly yields an anti-symmetrized version of the the scaling function $f(z)$.

\item  The derivation of universal expressions for $f(z)$ for systems of three different shapes (\ref{eq:fsquare},\ref{eq:ftr},\ref{eq:fhc}), based upon our standard definition of $b$.  Note that $f(z)$ is a subtle function to observe as it corresponds to finite-size corrections to $n(p)$.

\end{itemize}

\begin{table}[htpb]
\caption{Universality properties of various quantities related to $n(p)$.  A check in the first column means that the quantity depends upon the shape of the boundary of the system (with periodic b.\ c.); a check in the second column means that the quantity depends upon the lattice and percolation type (site or bond).  The final column shows the dependence on dimensionality, which applies to all of the quantities here. 
 }
\begin{center}
\begin{tabular}{lccc}
\hline \hline 
Quantity  &  Shape & Lattice & Dimensionality\cr
\hline
$\alpha,\nu \ldots$ &  & & \checkmark \cr
$\hat{\mathcal A}^\pm = b^{\alpha-2}{\mathcal A}^\pm$ &  & & \checkmark  \cr
$  A_0, B_0, C_0 $ &  &\checkmark & \checkmark  \cr
$b$ &  & \checkmark & \checkmark  \cr
$ f(z) $ & \checkmark & & \checkmark  \cr
$  A_1$, $b^{-1}B_1$, $b^{-2}C_1$, $R $ & \checkmark & & \checkmark  \cr
$B_1$, $C_1$ & \checkmark & \checkmark & \checkmark  \cr
\hline \hline 
\end{tabular}
\end{center}
\label{tab:universality}
\end{table}%


Future work is suggested to study $n(p)$ and $f(z)$ for different
lattices and boundary shapes, as well as the behavior in  higher dimensions.  Perhaps new
exact results for some of these quantities can also be found, such as $n(p_c)$ for site percolation
on the triangular lattice, where we found the precise value
(\ref{triangularsite}).  The dependence of $B_1/b$ and $C_1/b^2$ as a function of the system shape seems also interesting, since they are related to the scaling function $f(z)$.  

{\it Acknowledgments:}  IJ was supported under the Australian Research Council's Discovery Projects 
funding scheme by the grant DP140101110 and IJ's  computational work was undertaken with the assistance of resources and services from the National Computational Infrastructure (NCI), which is supported by the Australian Government.   The authors thank Peter Kleban for help in calculating the conformal excess number $A_1$ for the three different system shapes.


\bibliography{biblio.bib}

\end{document}